# Multigap superconductivity at extremely high temperature: a model for the case of pressurized $H_2S$

A. Bussmann-Holder[1], J. Köhler[1], A. Simon[1], M. Whangbo[2], A. Bianconi[3,4]

[1]*Max-Planck-Institute for Solid State Research, Heisenbergstr. 1, D-70569 Stuttgart, Germany*

[2]*Department of Chemistry, North Carolina State University, Raleigh, NC 27695-8204, USA*

[3]*RICMASS, Rome International Center for Materials Science Superstripes, Via dei Sabelli 119A, 00185 Rome, Italy,*

[4]*National Research Nuclear University, MEPhI, Kashirskoye sh. 31, Moscow 115409, Russia*

**Abstract:**

It is known that in pressurized $H_2S$ the complex electronic structure in the energy range of 200 meV near the chemical potential can be separated into two electronic components, the first characterized by steep bands with a high Fermi velocity and the second by flat bands with a vanishing Fermi velocity. Also the phonon modes interacting with electrons at the Fermi energy can be separated into two components: hard modes with high energy around 150 meV and soft modes with energies around 60 meV. Therefore we discuss here a multiband scenario in the standard BCS approximation where the effective BCS coupling coefficient is in the range 0.1-0.32. We consider a first (second) BCS condensate in the strong (weak) coupling regime 0.32 (0.15). We discuss different scenario segregated in different portions of the material. The results show the phenomenology of unconventional superconducting phases in this two-gap superconductivity scenario where there are two electronic components in two Fermi surface spots, the pairing is mediated by either by a soft or a hard phonon branch where the inter-band exchange term, also if small, plays a key role for the emergence of high temperature superconductivity in pressurized sulfur hydride.

After the discovery of high temperature superconductivity in cuprates, the search for materials with even higher transition temperatures $T_c$ was intensified, however, without success. Only recently the record $T_c$ of 165K obtained in cuprates was broken by hydrogen sulfide under ultrahigh pressure, which reaches a $T_c$ of 203 K. This new discovery was motivated by theoretical considerations where an enhancement of $T_c$ was predicted to take place in hydrogen containing compounds due to the light mass of hydrogen. Amazingly, the observed values of $T_c$>200K were rapidly classified as conventional, in the sense of BCS or Eliashberg theory whereas the first reports by Bednorz and Müller of $T_c$>30K immediately called for novel pairing interactions. Here we show that superconductivity in $H_2S$ cannot be accounted for by using standard approaches, but that several electronic bands with substantially weaker and stronger coupling strength, are involved in it where important interband interactions are essential in obtaining the high values of $T_c$ observed experimentally.

Already in 1968 Ashcroft predicted that high temperature superconductivity could be realized in metallic hydrogen [1]. These ideas have been followed over the years [2 – 6] and only recently concrete suggestions have been made that pressure induced metallization of dense $H_2S$ should be a high temperature superconductor [7,8]. Indeed this was realized in $H_2S$ where superconductivity was observed at 203K and at a pressure of 150GPa [9]. A substantial isotope effect on $T_c$ upon deuteration was reported [10]; the isotope effect varies with pressure and even exceeds the BCS value of 0.5. Since this observation strongly suggests an involvement of phonons in the pairing mechanism it was concluded that superconductivity in $H_2S$ is conventional. In the following we argue that this is not the case.

When high temperature superconductivity was discovered in cuprates [11], the $T_c$ value of 32K was immediately claimed to be incompatible with BCS or Eliashberg theory since the upper limit on $T_c$ was set around 28K [12, 13]. Meanwhile it is well accepted for this material class that the conventional mechanism does not work, however, consensus about a common pairing mechanism has not yet been achieved. It has been shown that multiple bands are involved in the pairing [14 – 16] and that interband interactions play a crucial role in the process of enhancing $T_c$ as well as in explaining the doping dependent isotope effect [17 – 19].

For $H_2S$ a similar approach is proposed, namely, that at least two bands are involved in the pairing mechanism with one band being rather localized in character whereas the other one is itinerant. This corresponds closely to the steep band / flat band scenario, [20]. The latter has been frequently discussed in the context of cuprate superconductivity where, however, not the coexistence of both has been emphasized but the crossover between them which corresponds to the transition from k-space pairing to real space pairing [21–25]. We invoke this scenario also for $H_2S$ since band structure calculations provide evidence that a flat band is located at the Fermi energy in coexistence with steep bands (Figure 1) quite analogous to MgB$_2$ [28].

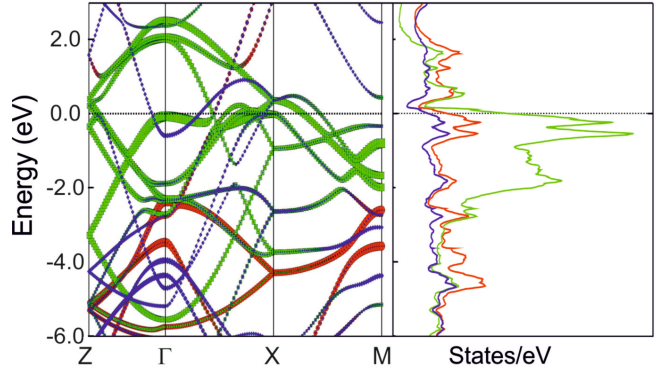

**Figure 1** Electronic structure calculated for the ultrahigh-pressure phase of H$_2$S with the perovskite structure (SH$^-$)(H$_3$S$^+$). a) Band dispersion relations with fat-band representations; the 3p states of the A site S atoms are shown in green, those of the B site S atoms in red, and the s/p states of the H atoms in blue. b) PDOS plots for the 3p states of the A site S atoms (green), the B site S atoms (red), and the s/p states of the H atoms (blue).

The Hamiltonian for this model is given by [29, 30]:

$$H = H_1 + H_2 + H_3 \tag{1}$$

$$H_1 = \sum_{k,\sigma}\left[\varepsilon_1(k)a^+_{k,\sigma}a_{k,\sigma} + \varepsilon_2(k)b^+_{k,\sigma}b_{k,\sigma}\right] \tag{1a}$$

$$H_2 = -\frac{1}{V}\sum_{k \neq g}[V_1 a^+_{k\uparrow}a^+_{-k\downarrow}a_{-g\downarrow}a_{g\uparrow} + V_2 b^+_{k\uparrow}b^+_{-k\downarrow}b_{-g\downarrow}b_{g\uparrow}] \tag{1b}$$

$$H_{12} = -\frac{1}{V}\sum_{k \neq g}[V_{12}(a^+_{k\uparrow}a^+_{-k\downarrow}b_{-g\downarrow}b_{g\uparrow} + b^+_{k\uparrow}b^+_{-k\downarrow}a_{-g\downarrow}a_{g\uparrow})] \tag{1c}$$

Here V is the volume and the terms in $H_i$ $(i = 1 - 3)$ are momentum k, g dependent. Electron creation and annihilation operators are denoted $a^+, a$ in band 1 and $b^+, b$ in band 2. The effective *intraband* pairing potentials are given by $V_1, V_2$, whereas $V_{12}$ stems from *interband* pair scattering. The band energies $\epsilon_i(k)$ reflect the flat band and the steep one and are correspondingly approximated by:

$\varepsilon_1 = const. = B;\ \varepsilon_2(k) = k^2/2m.$

Note that this approximation is equivalent to the coexistence of strong coupling flat and weak coupling steep bands.

After performing a Bogoliubov transformation the gap equations are explicitly obtained as:

$$\Delta_1 = \frac{V_1}{V} \sum_k \frac{\Delta_1}{\Omega_1(k)} \tanh\left[\frac{\Omega_1(k)}{2kT}\right] + \frac{V_{12}}{V} \sum_k \frac{\Delta_2}{\Omega_2(k)} \tanh\left[\frac{\Omega_2(k)}{2kT}\right] \qquad (2a)$$

$$\Delta_2 = \frac{V_2}{V} \sum_k \frac{\Delta_2}{\Omega_2(k)} \tanh\left[\frac{\Omega_2(k)}{2kT}\right] + \frac{V_{12}}{V} \sum_k \frac{\Delta_1}{\Omega_1(k)} \tanh\left[\frac{\Omega_1(k)}{2kT}\right] \qquad (2b)$$

with $\Omega_i(k) = \sqrt{\varepsilon_i(k)^2 + \Delta_i^2}$. These equations have to be solved simultaneously and self-consistently for each temperature T in order to derive the temperature dependence of the coupled gaps. The critical temperature $T_c$ is given by the condition $\Delta_1, \Delta_2 \rightarrow 0$ to yield:

$$\Delta_1 = \frac{V_1}{V} \sum_k \frac{\Delta_1}{\varepsilon_1(k)} \tanh\left[\frac{\varepsilon_1(k)}{2kT_c}\right] + \frac{V_{12}}{V} \sum_k \frac{\Delta_2}{\varepsilon_2(k)} \tanh\left[\frac{\varepsilon_2(k)}{2kT_c}\right] \qquad (3a)$$

$$\Delta_2 = \frac{V_2}{V} \sum_k \frac{\Delta_2}{\varepsilon_2(k)} \tanh\left[\frac{\varepsilon_2(k)}{2kT_c}\right] + \frac{V_{12}}{V} \sum_k \frac{\Delta_1}{\varepsilon_1(k)} \tanh\left[\frac{\varepsilon_1(k)}{2kT_c}\right] \qquad (3b)$$

Upon replacing the summations by integrals and introducing the density of states at the Fermi level $N_i(0)$, dimensionless coupling constants are defined as $\lambda_i = N_i(0)V_i$, $\lambda_{12} = \sqrt{N_1(0)N_2(0)}V_{12}$. With these definitions the above equations become:

$$1 = \lambda_1 \int_0^{\hbar\omega_1} \frac{1}{\varepsilon_1(k)} \tanh\left[\frac{\varepsilon_1(k)}{2kT_c}\right] d\varepsilon_1 + \lambda_{12} \int_0^{\hbar\omega_{1,2}} \frac{1}{\varepsilon_2(k)} \tanh\left[\frac{\varepsilon_2(k)}{2kT_c}\right] d\varepsilon_2 \qquad (4a)$$

$$1 = \lambda_2 \int_0^{\hbar\omega_2} \frac{1}{\varepsilon_2(k)} \tanh\left[\frac{\varepsilon_2(k)}{2kT_c}\right] d\varepsilon_2 + \lambda_{12} \int_0^{\hbar\omega_{1,2}} \frac{1}{\varepsilon_1(k)} \tanh\left[\frac{\varepsilon_1(k)}{2kT_c}\right] d\varepsilon_1 \qquad (4b)$$

By explicitly considering the band energies given above, eqs. 4a, b can be reformulated like:

$$1 = \lambda_1 \int_0^{\hbar\omega_1} \frac{1}{B} \tanh\left[\frac{B}{2kT_c}\right] d\varepsilon_1 + \lambda_{12} \int_0^{\hbar\omega_{1,2}} \frac{1}{k^2/2m} \tanh\left[\frac{k^2/2m}{2kT_c}\right] dk \qquad (5a)$$

$$1 = \lambda_2 \int_0^{\hbar\omega_2} \frac{1}{k^2/2m} \tanh\left[\frac{k^2/2m}{2kT_c}\right] dk + \lambda_{12} \int_0^{\hbar\omega_{1,2}} \frac{1}{B} \tanh\left[\frac{B}{2kT_c}\right] d\varepsilon_1 \qquad (5b)$$

The coupled integrals can be solved analytically only for certain limits of tanh(x): $\tanh(x) \rightarrow 1$, for x>>1, and $\tanh(x) \rightarrow x$ for x<<1. All intermediate cases have to be solved numerically, as discussed below.

Instead of using a parabolic dispersion for $\varepsilon_2(k)$ the BCS expression is used in the following which corresponds to an itinerant band character. This admits to solve the integrals:

$$1 = \lambda_1 \int_0^{\hbar\omega_1} \frac{1}{B} \tanh\left[\frac{B}{2kT_c}\right] d\varepsilon_1 + \lambda_{12} \int_0^{\hbar\omega_{1,2}} \frac{1}{\varepsilon_2(k)} \tanh\left[\frac{\varepsilon_2(k)}{2kT_c}\right] d\varepsilon_2 \qquad (6a)$$

$$1 = \lambda_2 \int_0^{\hbar\omega_2} \frac{1}{\varepsilon_2(k)} \tanh\left[\frac{\varepsilon_2(k)}{2kT_c}\right] d\varepsilon_2 + \lambda_{12} \int_0^{\hbar\omega_{1,2}} \frac{1}{B} \tanh\left[\frac{B}{2kT_c}\right] d\varepsilon_1 \qquad (6b)$$

$$1 = \frac{\lambda_1 \hbar\omega_1}{B} \tanh\left[\frac{B}{2kT_c}\right] + \lambda_{12} \ln\left[\frac{1.13\hbar\omega_{1,2}}{kT_c}\right] \qquad (7a)$$

$$1 = \lambda_2 \ln\left[\frac{1.13\hbar\omega_2}{kT_c}\right] + \frac{\lambda_{12}\hbar\omega_{1,2}}{B} \tanh\left[\frac{B}{2kT_c}\right] \qquad (7b)$$

The two approximate solutions are:

1. x<<1:

$$1 = \lambda_1 \frac{\hbar\omega_1}{2kT_c} + \lambda_{12} \ln\left[\frac{1.13\hbar\omega_{1,2}}{kT_c}\right] = \lambda_{12} \frac{\hbar\omega_{1,2}}{2kT_c} + \lambda_2 \ln\left[\frac{1.13\hbar\omega_2}{kT_c}\right] \qquad (8)$$

yielding an implicit relation for $T_c$.

2. x>>1:

$$1 = \frac{\lambda_1 \hbar\omega_1}{B} + \lambda_{12} \ln\left[\frac{1.13\hbar\omega_{1,2}}{kT_c}\right] = \lambda_{12} \frac{\hbar\omega_{1,2}}{B} + \lambda_2 \ln\left[\frac{1.13\hbar\omega_2}{kT_c}\right] \qquad (9)$$

which can be solved explicitly:

$$kT_c = 1.13\hbar\omega_2 \exp\left(\frac{-\lambda_2}{\lambda_{12}-\lambda_2}\right) + 1.13\hbar\omega_{1,2} \exp\left(\frac{-\lambda_{12}}{\lambda_2-\lambda_{12}}\right) + \exp\left(\frac{\lambda_{12}\hbar\omega_{12} - \lambda_1\hbar\omega_1}{B(\lambda_2-\lambda_{12})}\right) \qquad (10)$$

This is in contrast to the case where the phonon cutoff energies are identical, and the band dispersion remains undefined as discussed in the general two-band approach introduced by Suhl, Matthias and Walker (SMW) [30]:

$$kT_c = 1.13\hbar\omega \exp\left(-\frac{1}{\lambda}\right); \quad 1/\lambda = \frac{1}{2}\left[\lambda_1 + \lambda_2 \pm \sqrt{(\lambda_1-\lambda_2)^2 + 4\lambda_{12}^2}/[\lambda_1\lambda_2 - \lambda_{12}^2]\right] \qquad (11)$$

In the following the coupled gap equations are solved numerically using the assumption that a first strong-coupling band coexists with a second weak-coupling band. In this case one superconducting gap (strong-coupling band) is substantially larger than the second (weak-coupling band) one. The flat / steep band scenario is described here by two bands a flat one in a stronger coupling regime and the steep one in a weaker coupling regime. For both cases the enhancement of $T_c$ caused by the interband interaction is calculated as well as the temperature dependence of the related gaps and the isotope effects on $T_c$.

By choosing substantially different values of the cutoff frequencies for the two bands, namely, $\hbar\omega_1 = 115\ meV$ for the strong-coupling and $\hbar\omega_2 = 60\ meV$ for the weak-coupling band, keep $\lambda_1 = 0.32$ and $\lambda_2 = 0.14$ identical for all cases, $\varepsilon_1, \varepsilon_2$ are given via the self-consistency relations and as long as the zero temperature gaps are fixed as well.

Since the interband coupling is the decisive parameter which determines $T_c$ and the magnitude of the gaps, it is kept constant at a small value of $\lambda_{12} = 0.032$ in a first step. Using these three quantities as input, the intraband couplings $\lambda_i$ are predetermined and the gaps $\Delta_i$ (i=1, 2) have to be calculated self-consistently for each temperature T. The results for the temperature dependencies of $\Delta_i$ are shown in Figure 2 where the following three scenarios are compared to each other under the constraint that the interband couplings remain identical, $\lambda_{12} = \lambda_{21}$, the phonon energies are fixed and the zero temperature gaps are the same in order to make the comparison between the cases possible:

1. The SMW approach where the phonon cutoff energies are identical: black symbols and lines

2. A strong-coupling *flat* band combined with a weak-coupling *steep* (blue symbols and lines.

3. A *steep band* in the strong coupling regime combined with *flat* weak-coupling : green symbols and lines.

Note that for all three cases the intraband parameters and phonon cutoff frequencies are the same. At a first glance it is very striking that case 2 leads to an appreciable $T_c$ enhancement of almost a factor of 1.5 compared to the other alternatives. In addition, the temperature dependencies of the gaps is very different from those of the other scenarios, since an almost linear temperature-dependence is realized for temperatures up to 20K followed by a BCS type dependence for higher temperatures. The other cases considered above show the typical behavior expected for the SMW model and the BCS temperature dependence of both gaps (see inset to Figure 1) for the itinerant/flat band case. These very different temperature dependencies of the gaps for the three cases considered above can be taken as evidence for either of the scenarios to be realized when comparing with the experimental data. The gap to $T_c$ ratios are summarized in Table 1. While the ratios for the SMW and the itinerant / flat band cases yield the BCS value, this is substantially different for the flat / itinerant case, where it is considerably decreased.

**Table 1**: $T_c$ (in K), gap to $T_c$ ratios and average gap to $T_c$ ratios for the SMW, the flat band / itinerant and the itinerant / flat band models discussed in the text.

|  | SMW | Flat / itinerant | Itinerant / flat |
|---|---|---|---|
| $T_c$ (K) | 42 | 59 | 43 |
| $2\Delta_1/k_B T_c$ | 5.85 | 4.16 | 5.715 |
| $2\Delta_2/k_B T_c$ | 1.067 | 0.91 | 1.29 |
| $(\Delta_1+ \Delta_2)/k_B T_c$ | 3.459 | 2.535 | 3.52 |

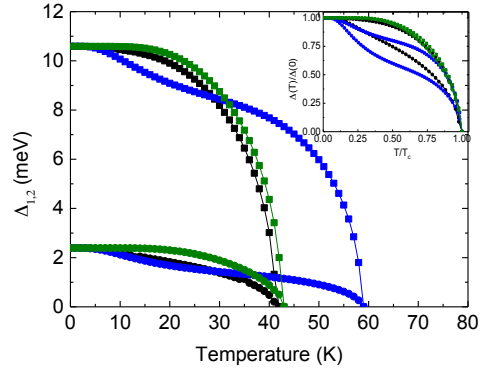

**Figure 2** Superconducting energy gaps $\Delta_1$, $\Delta_2$ as a function of temperature. Black symbols and lines refer to the SMW case, blue symbols correspond to the flat strong coupling band / itinerant weak coupling model, and green symbols and lines represent the itinerant strong coupling / flat weak coupling case.

An important consequence of the two-band, respectively the multiband approach is the enormous increase in $T_c$ caused by the interband interaction. This is even realized when the interband interaction is repulsive in the SMW model (Eq. 11) but not obvious in the two remaining cases where a transparent analytical formula for $T_c$ cannot be derived. Thus, the $T_c$-determining equations 6a, 6b have to be solved simultaneously and self-consistently. We consider only the two cases, namely, the flat band / itinerant and the itinerant / flat band, since the SMW model has already been studied in detail. The results are shown in Figure 3.

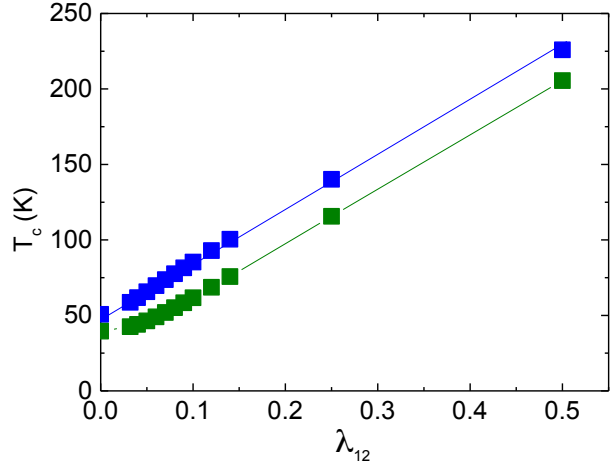

**Figure 3** $T_c$ as a function of the interband coupling $\lambda_{12}$ for the flat band / itinerant (blue) and the itinerant / flat band (green) case.

Note, that for both cases $T_c$ for $\lambda_{12} = 0$ is already rather large due to the choice of the high cutoff frequencies. With finite and increasing $\lambda_{12}$ $T_c$ grows rapidly and almost linearly to reach values of 200K for $\lambda_{12}$ between 0.4 and 0.5. The difference between the flat band / itinerant and itinerant / flat band case lies in the magnitude of $T_c$ which is systematically lower by approximately 20K for the latter case again due to the cutoff frequency which is less than half of the former case. The development of $T_c$ with increasing $\lambda_{12}$ is very different from the SMW model since in that approach a nonlinear increase in $T_c$ is observed [31]. This implies that the use of a flat band significantly modifies the development of $T_c$ with $\lambda_{12}$.

The multiband approach to superconductivity in $H_2S$ relates to the isotope effect which is positive and pressure dependent [10]. While its observation has been the origin to classify this compound as conventional [32 – 36], its pressure dependence [37,38] directly suggests an unconventional approach since BCS theory does not allow for a pressure dependent isotope exponent as long as the structure remains the same. In [32] the pressure induced structural changes have been explicitly taken into account, whereby the isotope coefficient can adopt different values for different transition temperatures. In the above approach two phonon frequencies are used which both can contribute to the isotope effect either due to the intraband or interband interactions (Eqs. 6a, 6b).

Here we discuss the isotope exponent in our scenario. Eqs. 6a and 6b have to be solved simultaneously and self-consistently to obtain $T_c$. Basically four possible sources for an isotope effect on $T_c$ are obvious from these equations, namely one related to the individual intraband interactions, or one stemming from the interband terms. Of course also combinations of these interactions are possible, but not discussed here for simplicity. It is, however, important to emphasize that any deviation of the isotope exponent from the BCS value is caused by the use of two or more bands as pairing sources.

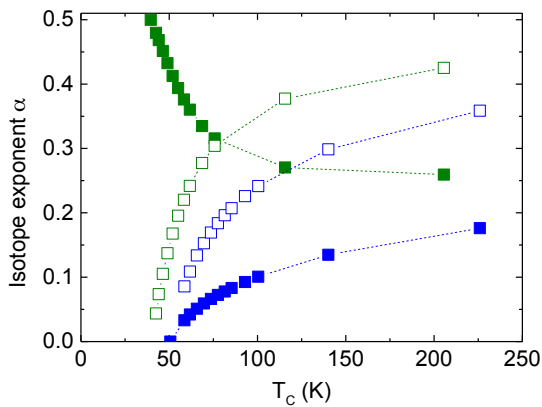

**Figure 4** Isotope exponent α as a function of $T_c$ for the cases: intraband flat band / steep (full blue symbols), intraband steep / flat band (full green symbols). The green open symbols refer to interband steep / flat band, the open blue symbols to interband flat band / steep, respectively.

All four resulting possibilities have been evaluated numerically and are shown in Figure 4. Interestingly, the interband interaction related isotope effects and the one stemming from the strong coupling band / weak coupling scenario increase with increasing $T_c$ to almost reach the BCS value of 0.5 for the interband derived isotope exponents and converge to 0.2 for the intraband one.

A distinctly different behavior is, however, seen for the intraband weak/ strong coupling band case where α decreases with increasing $T_c$ to approach a value of 0.25 at the maximum $T_c$. This trend corresponds qualitatively to the one seen experimentally. Again strong deviations from the SMW model are apparent where α in the case of interband related isotope effect increases rapidly to values exceeding the BCS value, whereas the two interband related isotope effects merge from 0.5, 0 to 0.25 at high $T_c$.

It has been speculated recently that the pressure dependent isotope effect must be related to strong anharmonicity stemming from the hydrogen motion [33]. This is, however, in contrast to early measured isotope effects in PdH, where a sign reversal of α takes place upon deuteration related to anharmonicity [39 – 41]. Such a sign reversal is clearly absent in $H_2S$ suggesting that anharmonicity of hydrogen motion is not the origin of the isotope effect.

From the results described above, we conclude that it is possible that the multi band scenario of strong – weak coupling in steep and flat bands is realized in $H_2S$. This not only leads to high values of $T_c$ for moderate $\lambda_{12}$, but also it predicts a pressure dependence of α.